\documentclass[12pt]{article}
\usepackage{graphicx}

\def\lsim{\raise0.3ex\hbox{$<$\kern-0.75em\raise-1.1ex\hbox{$\sim$}}}
\def\gsim{\raise0.3ex\hbox{$>$\kern-0.75em\raise-1.1ex\hbox{$\sim$}}}
\def\noi{\noindent}

\def\beq{\begin{equation}}   \def\eeq{\end{equation}}
\def\bea{\begin{eqnarray}}  \def\eea{\end{eqnarray}}

\def\noi{\noindent}

\begin{document}
\pagestyle{empty}
{\hfill LPT-Orsay 04-49}
\vskip 2 truecm

\begin{center}

{\bf \large Real and Virtual Photoproduction of }\par \vskip 3 truemm 
{\bf \large Large-p$_{\bot}$ Particles at NLO}
\vskip 1 truecm

{\bf M. Fontannaz}\\
Laboratoire de Physique Th\'eorique (UMR 8627 - CNRS)\\
         Universit\'e Paris XI, b\^atiment 210, 91405 Orsay cedex, France

\end{center}

\vskip 2 truecm
 
\begin{abstract}
In the first part of this paper we assess the possibility of observing
the gluon distribution in a real photon by measuring the
photoproduction cross section of large-$p_{\bot}$ photons. In the
second part we calculate the virtual photoproduction of
large-$p_{\bot}$ forward $\pi^0$. The theoretical results are compared
with data and with BFKL-inspired predictions. These studies are done at
the NLO approximation.
\end{abstract}

\par \vskip 3 truecm

\begin{center}
   {\it \large To appear in the proceedings of the\\ 7th DESY Workshop on
Elementary Particle Theory\\ ``Loops and Legs in Quantum Field Theory'',\\
Zinnowitz, Germany, April 2004.}
\end{center}

%\flushleft

\newpage

\pagestyle{plain}

\section{Introduction}
The real and virtual photoproduction of large-$p_{\bot}$ particles
gives rise to interesting tests of the QCD dynamics and enables to
measure the parton distribution and fragmentation functions \cite{1r}.
In particular it offers a unique opportunity to measure the parton
distributions in the (real or virtual) photon. In this paper I shall
concentrate on this feature of photoproduction reactions. In the first
part, I assess the possibility to determine the gluon distribution in a
real photon by measuring the photoproduction of large-$p_{\bot}$
photons, and in the second part, I study the importance of the virtual
photon structure function in the forward (along the initial proton
direction) production of large -$p_{\bot}$ $\pi^0$. This latter
reaction is also interesting from the point of view of the underlying
QCD dynamics. In fact it has been proposed by Mueller \cite{2r} in
order to study the importance of the BFKL contribution to the forward
cross section.

\section{The gluon distribution in the real photon} Over the past
years, the ZEUS \cite{3r,4r} and H1 \cite{5r} collaborations at HERA
have been able to observe the photoproduction of large$-p_{\bot}$
photons, and the comparisons of data with existing NLO QCD prdictions
\cite{6r,7r,8r,9r,10r} appear successful. In photoproduction reactions,
a quasi-real photon, emitted at small angle from the electron,
interacts with a parton from the proton. The photon can either
participate directly in the hard scattering or be resolved into a
partonic system, in which case the parton stemming from the photon
takes part in the hard interaction. Therefore photoproduction is a
privileged reaction to measure or constrain the parton distributions in
the photon and in the proton. Here I will put the emphasize on the
gluon distribution in the real photon, a distribution which is hardly
known \cite{11r}.\par

The interest of the reaction $\gamma + p \to \gamma + jet + X$ comes
from the fact that the final photon offers a clear experimental signal.
On the theoretical side, this cross section is hardly sensitive to the
factorization and renormalization scales, a fact which should allow us
an accurate determination of the parton distributions.\par

As observables which serve to reconstruct the longitudinal momentum
fraction of the parton stemming from the photon, it is common to use
\beq \label{1e} x_{obs}^{\gamma} = {p_T^{\gamma} \ e^{-\eta^{\gamma}} +
E_T^{jet} \ e^{-\eta^{jet}} \over 2 E^{\gamma}}\ . \eeq

\noi However, as the measurement of $E_T^{jet}$ can be a substantial
source of systematic errors at low $E_T$ values, we propose a slightly
different variable which does not depend on $E_T^{jet}$, \beq
\label{2e} x_{LL}^{\gamma} = {p_T^{\gamma} ( e^{-\eta^{\gamma}} +
e^{-\eta^{jet}}) \over 2 E^{\gamma}}\ . \eeq

\noi At leading order, for the non-fragmentation contribution, the
variables $x_{obs}$ and $x_{LL}$ coincide, and they are also equal to
the ``true'' partonic longitudinal momentum fraction, i.e. the argument
of the parton distribution function. At NLO, the real corrections
involve 3 partons in the final state (with transverse momenta $p_{T3}$,
$p_{T4}$, $p_{T5}$), one of which -- say parton 5 -- is unobserved.
Therefore, $d\sigma/x_{obs}$ and $d\sigma/x_{LL}$ will be different at
NLO.\par

Before presenting numerical results, let us further discuss the problem
of the cuts on the photon and jet transverse momenta. It is well known
that symmetrical cuts on the minimum transverse energies of the photon
and of the jet should be avoided as they amount to including a region
where the fixed order perturbative calculation shows infrared
sensitivity. As explained in detail in \cite{11r}, the problem stems
from terms $\sim log^2(|1 - p_T^{\gamma}/E_{T,{min}}^{jet}|)$ which
become large as $p_T^{\gamma}$ approaches $E_{T,{min}}^{jet}$, the
lower cut on the jet transverse energy. Therefore the partonic NLO
cross section has a singular behaviour at $p_T^{\gamma} =
E_{T,{min}}^{jet}$. Obviously, $d\sigma/dp_T^{\gamma}$ will not exhibit
a problem as long as $E_{T,min}^{jet} < p_{T,min}^{\gamma}$ since the
critical point $p_T^{\gamma} = E_{T,min}^{jet}$ will not be reached in
this case. \par

On the other hand, one often would like to have a more inclusive cross
section such as $d\sigma/d\eta^{\gamma}$, obtained by integrating the
differential cross section over $p_T^{\gamma}$ and $E_T^{jet}$. In this
case one should not choose $p_{T,min}^{\gamma} = E_{T,{min}}^{jet}$ as
this amounts to integrating the spectrum over the $\log^2 ((1 -
p_{\bot}^{\gamma}/E_{\bot , {\rm min}}^{jet}|)$ contribution. As a
result, the theoretical prediction although being finite, is infrared
sensitive as a consequence of choosing symmetrical cuts. This point has
been discussed in detail in ref. \cite{11r}.\par

\begin{figure}[!]
\includegraphics[width=3.2in,height=3.2in]{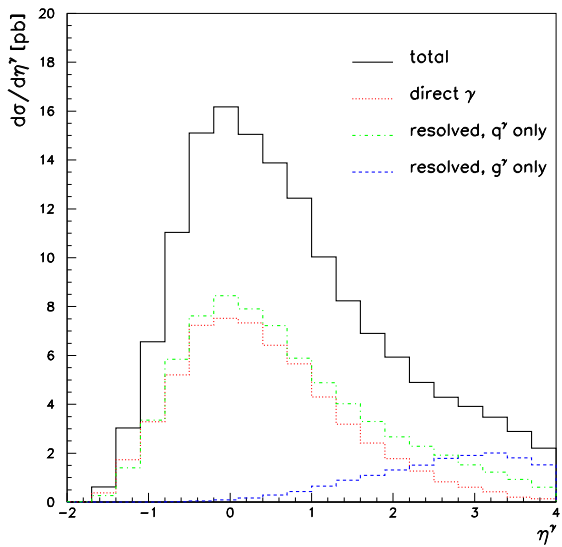}
\centering
\caption{Magnitude of different subprocesses over the full photon
rapidity range. The jet rapidities have been integrated over $- 2 <
\eta^{\rm jet} <4$, and $E_T^{\rm jet} > 5$~GeV, $p_T^{\gamma} >
6$~GeV.} \label{fig:1} \end{figure}

Our studies are based on the program EPHOX \cite{12r}, which is a  NLO
partonic Monte Carlo event generator. Unless stated otherwise, we use
the following input for our numerical results~: A center of mass energy
$\sqrt{s} = 318$~GeV with $E_e = 27.5$~GeV and $E_p = 920$~GeV is used.
The cuts on the minimum transverse energies of photon and jet are
$E_T^{\rm jet} > 5$~GeV, $p_T^{\gamma} > 6$~GeV. The rapidities have
been integrated over in the domain $- 2 \leq \eta^{\gamma}, \eta^{\rm
jet} \leq 4$ unless stated otherwise. For the parton distributions in
the proton we take the MRST01 parametrization, for the photon we use
AFG04 \cite{13r} distribution functions and BFG \cite{14r}
fragmentation functions. We take $n_f = 4$~flavours, and for $\alpha_s
(\mu )$ we use an exact solution of the two-loop renormalization group
equation, and not an expansion in $\log (\mu/\Lambda )$. The default
scale choice is $M = M_F = \mu = p_T^{\gamma}$. Jets are defined using
the $k_T$-algorithm. The rapidities refer to the $ep$ laboratory frame,
with the HERA convention that the proton is moving towards positive
rapidity. One must also note that the final photon verifies an
isolation criterion \cite{11r}. \par

The main result of the study is shown in Fig.~1 which details the
various contributions to the cross section.

%\begin{figure}[!]
%\includegraphics[width=3.2in,height=3.2in]{fig1.eps}
%\centering
%\caption{Magnitude of different subprocesses over the full photon
%rapidity range. The jet rapidities have been integrated over $- 2 <
%\eta^{\rm jet} <4$, and $E_T^{\rm jet} > 5$~GeV, $p_T^{\gamma} >
%6$~GeV.} %\label{fig:1} %\end{figure}

The gluon distribution $g^{\gamma}(x^{\gamma}, Q^2)$ only contributes
at small values of $x^{\gamma}$, corresponding to large values of
$\eta^{\gamma}$, and we shall try, by various cuts, to enhance the
relative contribution of this component. Cuts in the photon and jet
rapidities are quite effective to enhance the gluon in the photon, as
shown in Fig.~2.

\begin{figure}[!]
\includegraphics[width=3.1in,height=4.0in]{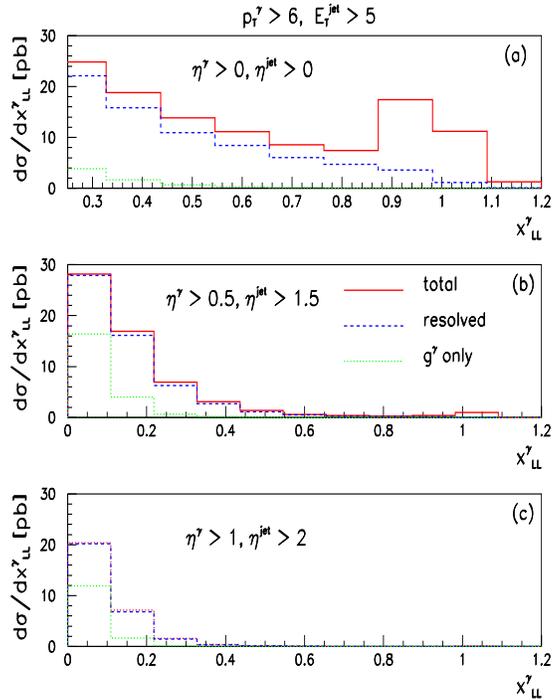}
\centering
\caption{Effect of
rapidity cuts to enhance the contribution from the gluon in the
photon.} \label{fig:2} \end{figure}

\begin{figure}[!]
\includegraphics[width=3.1in,height=4.0in]{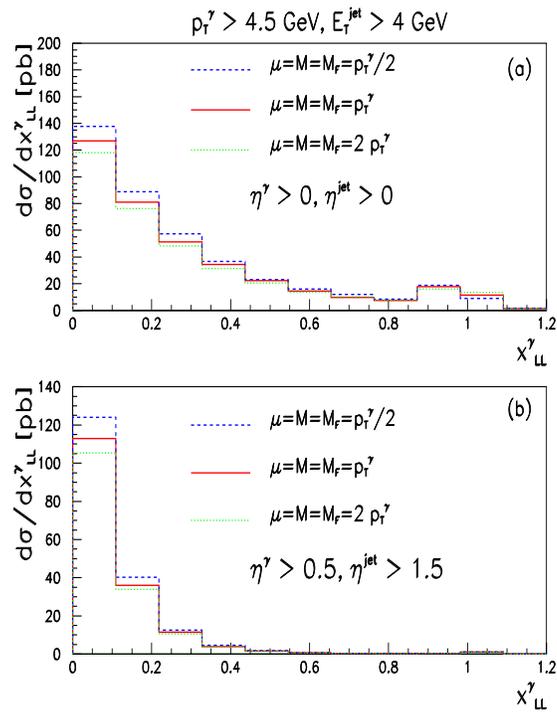}
\centering
\caption{Scale
dependence of $d\sigma/dx_{LL}^{\gamma}$ in the presence of forward
rapidity cuts.} \label{fig:3} \end{figure}

Note that the lower cuts on the transverse momenta are rather large,
$p_T^{\gamma} > 6$~GeV, $E_T^{\rm jet} > 5$~GeV. One can increase the
cross section by choosing lower $p_T$ cuts, as shown in Fig.~3. This
figure also shows the scale dependence of $d/\sigma/dx_{LL}^{\gamma}$
in the presence of the cuts $\eta^{\gamma} > 0$, $\eta^{jet} > 0$
respectively $\eta^{\gamma} > 0.5$, $\eta^{jet} > 1.5$. The behaviour
of the cross section $d\sigma/dx_{Ll}^{\gamma}$, varies by $\pm 8$~\%
under the scale changes. One must keep in mind that the distribution
$g^{\gamma}$ is poorly known and that a determination of the latter
with an accuracy of $\pm 10$~\% would be quite welcome.

\section{The forward leptoproduction of large-$p_{\bot}$ $\pi^0$} This
reaction has been put forward by Mueller \cite{2r} to observe the BFKL
dynamics. In fact in the collision between the virtual photon of
virtuality $Q^2 = |q^2|$ and a gluon from the incident proton, we can
have the reaction $\gamma^*(\to q\overline{q}) + g \to q \overline{q} +
ng$ with $n$ gluons emitted in the final state by a ladder of gluons
exchanged between the $q\overline{q}$-pair and the initial gluon. In
the configuration in which the large$-p_{\bot}$ forward $\pi^0$ is a
fragment of the first (closest to the proton) gluon, we have a gluonic
ladder, starting at virtuality $p_{\bot}^2$ and ending at the
virtuality $Q^2$. When $p_{\bot}^2 \sim Q^2$, there is no room for a
DGLAP evolution. However, when $x_{B_j} = Q^2/2P\cdot q$ is small, the
$\ell n x_{Bj}$ terms generated by the latter can be resumed by the
BFKL equation \cite{15newr}~; this leads to a large contribution to the
forward cross section if we follow the estimations of ref.
\cite{15r}.\par

In this work we follow another way and calculate the electroproduction
cross section at the NLO approximation. The cross section is the sum of
a direct term and of a resolved term. For both of them, we calculate HO
corrections the parton distributions in the virtual photon also are
calculated at the NLO approximation \cite{16r,17r}. This cross section
has been measured by the H1 collaboration \cite{18r,19r} and we here
present results for $d\sigma/dx_{B_j}$ \cite{18r} which verifies the
following constraints. In the laboratory system a $\pi^0$ is observed
in the forward direction with $5^{\circ} \leq \theta_{\pi^0} \leq
25^{\circ}$~; the laboratory momentum of the pion is constrained by
$x_{\pi^0} = E_{\pi^0}/E_P \geq .01$ and an extra cut is put on the
$\pi^0$ transverse momentum in the $\gamma^*-p$ center of mass system~:
$p_{\bot \pi^0}^* >$~2.5 GeV. The inelasticity $y = Q^2/x_{B_j}S$ is
restricted to the range $.1 < y <.6$, where $S = (p_e + p_P)^2 =
(300$~GeV)$^2$. We use $\Lambda_{\overline{MS}}^{(4)} = 300$~MeV and
the fragmentation functions of ref. \cite{20r}. All the factorization
and renormalization scales are taken equal to $Q^2 +
E_{\bot}^{\pi^{0^2}}$. The direct HO corrections from which the (lowest
order) resolved part is subtracted are called HO$_s$.\par
\begin{figure}[htb]
\includegraphics[width=3.1in,height=3.1in]{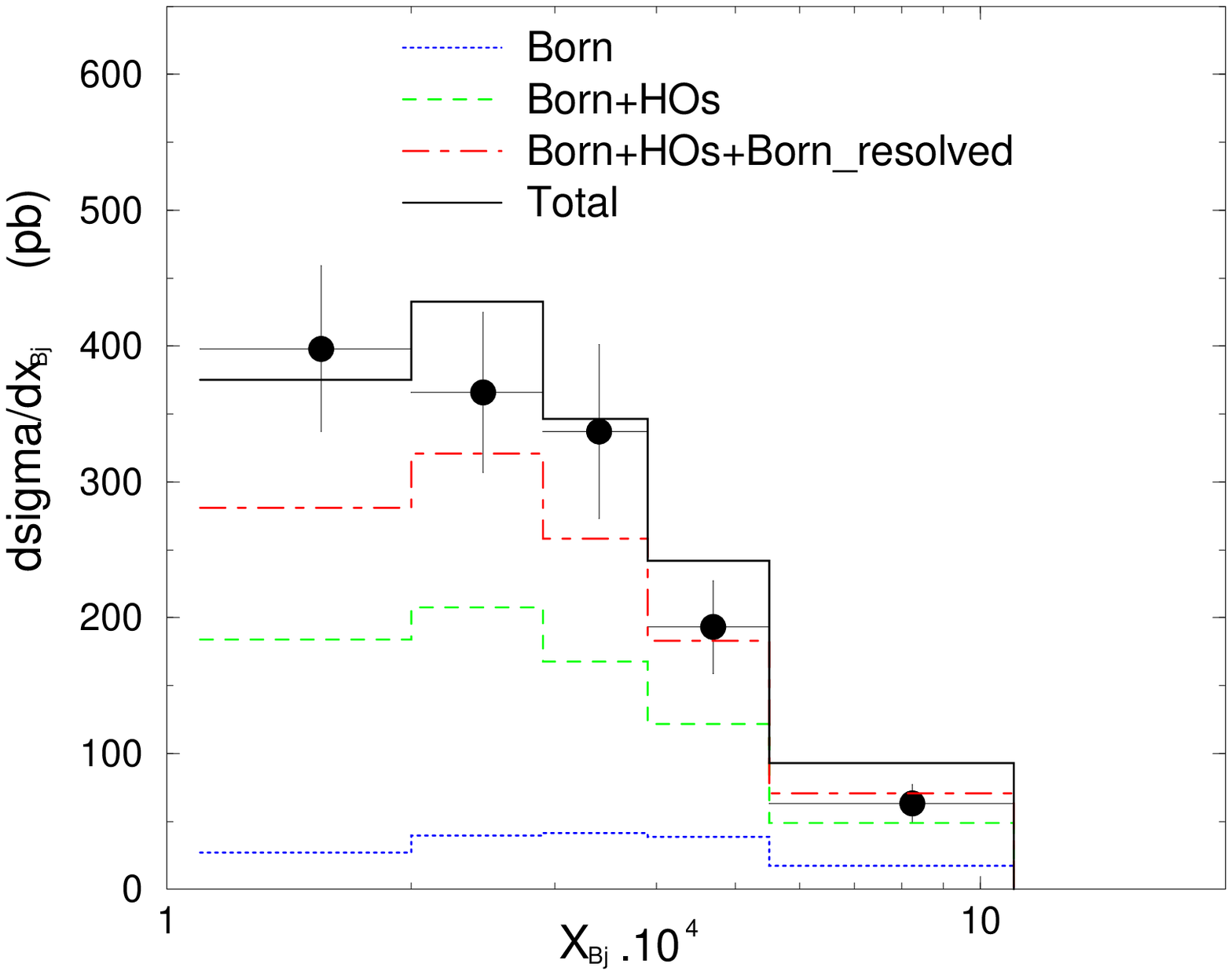}
\centering
\caption{}
\label{fig:4} \end{figure}

In Fig.~4 we notice the importance of the HO$_s$ corrections to the
direct Born cross section. These large corrections come from the
subprocesses $\gamma^* + g \to q + \overline{q} + g$ and $\gamma^* + q
\to q + \overline{q} + q$ that have a gluon exchanged in the
$t$-channel. We also notice the importance of the resolved contribution
(approximately one half of the total cross section) with large HO
corrections~: HO (resolved)/Born (resolved) $\simeq 1$. All these large
corrections are associated to small values of $E_{\bot}$ due to the
small cut $p_{\bot \pi^0}^* > 2.5$~GeV. As a consequence the total
cross section is quite sensitive to variation of the renormalization
scale $\mu$ \cite{16r,17r}. For the choice $\mu^2 = Q^2 + E_{\bot
\pi^0}^2$ we obtain a good agreement between data and theory with
little room left for a BFKL contribution as estimated in ref.
\cite{15r}.


\begin{thebibliography}{9} \bibitem{1r} M. Klasen, Rev. Mod. Phys. 74
(2002) 1221. \bibitem{2r} A. H. Mueller, Nucl. Phys. (Proc. Suppl.)
B18c (1990) 125. \bibitem{3r} J. Breitweg et al. [ZEUS Collaboration],
Phys. Lett. B472 (2000) 175 [hep-ex/9910045]~; \\ J. Breitweg et al.
[ZEUS Collaboration], Phys. Lett. B413 (1997) 201 [hep-ex/9708038].
\bibitem{4r} S. Chekanov et al. [ZEUS Collaboration], Phys. Lett. B511
(2001) 19 [hep-ex/0104001]. \bibitem{5r} H1 Collaboration, submitted to
the Int. Europhysics Conference on High Energy Physics, EPS03, July
2003, Aachen (Abstract 093), and to the XXI Int. Symposium on Lepton
and Photon Interactions, LP03, August 2003, Fermilab. \bibitem{6r} L.
E. Gordon, Phys. Rev. D57 (1998) 253 [hep-ph/97075464]. \bibitem{7r} M.
Fontannaz, J. P. Guillet and G. Heinrich, Eur. Phys. J. C21 (2001) 303
[hep-ph/0105121]. \bibitem{8r} M. Fontannaz, J. P. Guillet and G.
Heinrich, Eur. Phys. J C22 (2001) 303 [hep-ph/0107262]. \bibitem{9r} M.
Krawczyk and A. Zembrzuski, Phys. Rev. D64 (2001) 114017
[hep-ph/0105166]. \bibitem{10r} A. Zembrzuski and M. Krawczyk,
hep-ph/0309308. \bibitem{11r} M. Fontannaz and G. Heinrich, Eur. Phys.
J. C34 (2004) 191. \bibitem{12r} The program together with detailed
documentation is available at
http:$\setminus\setminus$www.lapp.in2p3.fr/lapth/PHOX\_
FAMILY/main.html. \bibitem{13r} P. Aurenche, J. P. Guillet and M.
Fontannaz, Z. Phys. C64 (1994) 621~;\\ P. Aurenche, J. P. Guillet and
M. Fontannaz, new version of AFG, publication in preparation.
\bibitem{14r} L. Bourhis, M. Fontannaz and J. P. Guillet, Eur. Phys. J.
C2 (1998) 529 \bibitem{15newr} V. S. Fadin, E. A. Kuraev, L. N.
Lipatov, Sov. Phys. JETP 44 (1976) 199. \\ Y. Y. Balitsky, L. N.
Lipatov, Sov. J. Nucl. Phys. 28 (1978) 822. \bibitem{15r} J.
Kwiecinski, A. D. Martin, J. Outhwaite, Eur. Phys. J. C9 (1999) 611.
\bibitem{16r} P. Aurenche, Rahul Basu, M. Fontannaz, R. Godbole, Eur.
Phys. J. C34 (2005) 277. \bibitem{17r} M. Fontannaz, in preparation,
preprint LPT-ORSAY-04-48. \bibitem{18r} H1 collaboration, C. Adloff et
al., Phys. Lett. B462 (1999) 440. \bibitem{19r} H1 Collaboration, A.
Aktas et al., DESY 04-051, hep-ex/0404009. \bibitem{20r} B. A. Kniehl,
G. Kramer, B. P\"otter, Nucl. Phys. B582 (2000) 514
\end{thebibliography}
\end{document}